\documentclass[12pt,a4paper]{article}
\usepackage[utf8]{inputenc}
\usepackage{a4wide}
\usepackage{graphicx}
\usepackage{amsmath}
\allowdisplaybreaks[1] 
\usepackage{hyperref}

\newcommand{\NN}{\ensuremath{\mathcal{N}}}
\newcommand{\NNc}{\ensuremath{\mathcal{N}^c}}
\newcommand{\NNt}{\ensuremath{\widetilde{\mathcal{N}^c}}}

\begin{document}

\begin{titlepage}
\begin{flushright}
LU TP 17-30\\
arXiv:1710.04383 [hep-ph]\\
revised December 2017
\end{flushright}
\vfill
\begin{center}
{\Large\bf Chiral Perturbation Theory for\\[2mm] Neutron-antineutron Oscillations}
\vfill
{\bf Johan Bijnens and Erik Kofoed$^\dagger$}\\[0.3cm]
{Department of Astronomy and Theoretical Physics, Lund University,\\
S\"olvegatan 14A, SE 223-62 Lund, Sweden}
\end{center}
\vfill
\begin{abstract}
We construct the Chiral Perturbation Theory operators for neutron-antineutron
oscillations and use these to estimate chiral and finite volume corrections
at one-loop order.
\end{abstract}
\vfill
{\footnotesize\noindent $^\dagger$ Present address: Narodowe Centrum Bada\'{n} J\c{a}drowych, Ho\.{z}a 69, 00-681 Warszawa, Poland}
\end{titlepage}

\section{Introduction}

The baryon asymmetry of the universe is one of the open problems in particle
physics. One possible solution is to have $B-L$ violation as exemplified in
$\Delta B=2$ transitions and in particular neutron-antineutron oscillations.
This has been suggested long ago, see e.g. \cite{Glashow:1979nm,Mohapatra:1980qe,Kuo:1980ew,Chang:1980ey,Chetyrkin:1980he}. 
Recent reviews are \cite{Mohapatra:2009wp,Phillips:2014fgb}. 
$\Delta B=2$ transitions require a six-quark operator. These were classified in
\cite{Rao:1982gt,Rao:1983sd,Caswell:1982qs}. To obtain predictions of a
particular model the coefficients of these operators need to be evolved to a
low scale and then the matrix elements computed. This running is known to
two-loop order \cite{Buchoff:2015qwa}. We will also use the notation of the
operators used in that reference. In the past these matrix elements were
estimated using models but now the first lattice calculations have appeared
\cite{Buchoff:2012bm,Syritsyn:2016ijx}. These can be done at different quark
masses from the physical ones and are necessarily at finite volume. Chiral
Perturbation Theory (ChPT) allows to do estimate both of these effects.

The bounds on the mean oscillation time $\tau$ are $\tau > 8.6~10^{7}$~s from
free neutrons \cite{BaldoCeolin:1994jz} and $\tau > 2.7~10^{8}$~s from bound
neutrons \cite{Abe:2011ky}. The reason why the bound from bound neutrons is
much lower than those for proton decay is that the antineutron inside nuclei
is far off-shell, see e.g. \cite{Gal:1999hx} for a clear explanation.
For the same reason, strong magnetic shielding is needed for the free neutron
experiments. A new free neutron experiment is proposed for ESS in Lund \cite{Milstead:2015toa} so a better estimate of the matrix elements will be very useful
to put limits on $\Delta B=2$ effects in theories beyond the Standard Model.

In this paper we construct the ChPT equivalents of the six-quark operators
of \cite{Buchoff:2015qwa} and use these then to calculate the chiral and finite
volume corrections in the isospin limit. The finite volume corrections are
found to be small for $m_\pi L>4$ for the physical pion mass but chiral
extrapolations can be substantial already for pion masses of order 200~MeV.

In Sect.~\ref{sec:quarks} we discuss shortly the quark operators
of \cite{Buchoff:2015qwa} and their chiral representation. Sect.~\ref{sec:ChPT}
discusses the ChPT aspects. The main new result is the construction of the
ChPT operators for neutron-antineutron transitions. This is done using the
spurion technique. In Sect.~\ref{sec:analytical} we calculate the one-loop
corrections in ChPT to the matrix elements and in Sect.~\ref{sec:numerical}
we give some numerical results. Our main conclusions are given in
Sect.~\ref{sec:conclusions}. App.~\ref{app:group} recalls some $SU(2)$
identities used heavily in deriving the ChPT operators and the needed integrals
are discussed in App.~\ref{app:integrals}.

Preliminary results of this work were presented in the master
thesis \cite{masterthesis} and at Lattice 2017 \cite{Lattice2017}. Related work
is in progress by Oosterhof et al. \cite{Oosterhof}.

\section{Quark operators and chiral properties}
\label{sec:quarks}

The operator structure needed for $n\bar n$-transitions contains six quark
fields $dddduu$ where under the chiral symmetry group $SU(2)_L\times SU(2)_r$
each quark field can be in a left- or right-handed doublet.
The operators were classified in \cite{Rao:1982gt,Rao:1983sd,Caswell:1982qs}
and rewritten in a basis that shows the chiral properties
in~\cite{Buchoff:2015qwa}. It was found that there are 14 operators
that have six types of representations under the chiral group.
There are three $(1_L,3_R)$, one $(1_L,7_R)$ and three $(5_L,3_R)$ operators,
as well as their parity conjugates. The chiral loop corrections for the
parity-conjugates are the same since the strong interactions are invariant
under parity.

If we assume isospin conservation, only an $I=1$ operator can contribute
to $n\bar n$-transitions. So only the $I=1$ projection of the different
$(5_L,3_R)$ and $(3_L,5_R)$ operators contributes, this explains why the loop
contributions for all those operators are the same, in fact one can show that
the operators $P_5,P_6,P_7$ (and similarly $Q_5,Q_6,Q_7$) are related by
isospin.
The $(1_L,7_R)$ and $(7_L,1_R)$ operators do not contribute in the isospin
limit. The operators are summarized in Tab.~\ref{tab:operators}.
\begin{table}
\centerline{
\begin{tabular}{|cll|cl|}
\hline
Chiral  & $\#$operators &Chiral & Spurion & $\#$operators \\
\hline
$(3_L,1_R)$ & 3: $P_1,P_2,P_3$ & $\theta_i^{i_L j_L}~(i=1,2,3)$        &$(1_L,3_R)$ & 3: $Q_1,Q_2,Q_3$\\
$(3_L,5_R)$ & 3: $P_5,P_6,P_7$ & $\theta_i^{i_L j_L k_R l_R m_R n_R}~(i=4,5,6)$        &$(3_R,5_L)$ & 3: $Q_5,Q_6,Q_7$\\
$(7_L,1_R)$ & 1: $P_4$         & $\theta_4^{i_L j_L k_L l_L m_L n_L}$        &$(1_L,7_R)$ & 1: $Q_4$\\
\hline
\end{tabular}}
\caption{\label{tab:operators} The chiral representations of the dimension-9
six-quark operators as listed in \cite{Buchoff:2015qwa} as well as the
corresponding spurions. The indices on the spurions are $SU(2)_L\times SU(2)_R$ upper doublet, fully symmetrized in the indices of the same type.}
\end{table}

We can add spurion fields transforming under $G_\chi=SU(2)_L\times SU(2)_R$
such that
the combination of quark-operators with chiral flavour indices and the spurions
is invariant under $G_\chi$. These will be used to construct the operators in
ChPT. There is a corresponding set for the opposite parity operators $Q_i$.

\section{Chiral perturbation theory}
\label{sec:ChPT}

We work in two-flavour ChPT and we use the heavy-baryon
formalism~\cite{Jenkins:1990jv} (HBCHPT),
a review and introduction is~\cite{Bernard:1995dp}.
The notation we use can be found in ~\cite{Bernard:1995dp}
or \cite{Ecker:1995rk}. The lowest order meson Lagrangian is
\begin{align}
\mathcal{L}_2 =\,& \frac{F^2}{4}\left\langle u_\mu u^\mu +\chi_+\right\rangle\,,
 & u_\mu =\,& i\left[u^\dagger (\partial_\mu-i r_\mu)u-u(\partial_\mu-i l_\mu)u^\dagger\right]\,,
\nonumber\\
\chi =\,& 2 B\left(s+i p\right)\,, &  \Gamma_\mu =\,& \frac{1}{2}\left[u^\dagger (\partial_\mu-i r_\mu)u-u(\partial_\mu-i l_\mu)u^\dagger\right]\,,
\nonumber\\
\chi_\pm =\,& u^\dagger \chi u^\dagger \pm u \chi^\dagger u\,,
 & \left\langle A \right\rangle \equiv\,& \mathrm{tr}(A)\,.
\end{align}
$u$ is a $2\times 2$ unitary matrix that contains the pion fields $\pi^a$
via $u=\exp(\pi^a \tau^a/(2F))$, with $\tau^a$ the Pauli matrices.
$B,F$ are the two lowest-order (LO)
low-energy constants (LECs). The $2 \times 2$ matrices $s,p,l_\mu,r_\mu$ are the
usual ChPT external fields.

Under a chiral transformation $g_L,g_R$ the
objects above transform as
\begin{align}
u\to\,& g_L u h^\dagger \equiv h u g_L^\dagger\,, &
  u_\mu\to\,& h u_\mu h^\dagger\,,
& \chi\to\,& g_R\chi g_L^\dagger\,,
\nonumber\\
\chi_\pm\to\,& h\chi_\pm h^\dagger\,, & U=u^2\to\,& g_R U g_L^\dagger\,.
\end{align}
The first equation is the definition of the compensator transformation $h$ which
depends on $u,g_L,g_R$. The last one defines $U$.

\begin{figure}
\centerline{
\includegraphics[width=0.4\textwidth]{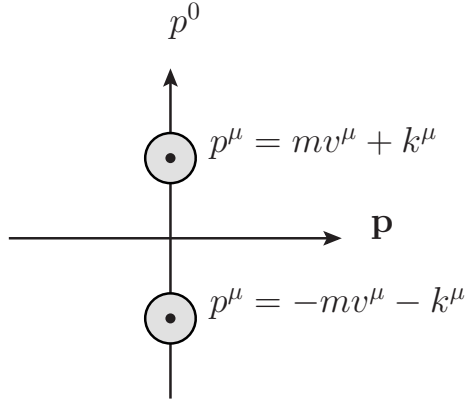}}
\caption{\label{fig:velocities}A pictorial representation of the velocity regions
relevant for projection on the nucleon and antinucleon HBCHPT fields.}
\end{figure}

Nucleons in a relativistic normalization can be included via a doublet field
$\Psi$ at LO as~\cite{Gasser:1987rb}
\begin{align}
\Psi =\,&\left(\begin{array}{c}p\\n
  \end{array}
\right)\,, & \psi\to\,& h \Psi\,,
\nonumber\\
\mathcal{L}_R =\,&\overline\Psi \left(iD_\mu \gamma^\mu - m+\frac{g_A}{2}u_\mu \gamma^\mu \gamma_5\right)\Psi\,, & D_\mu \equiv\,& \partial_\mu+\Gamma_\mu\,.
\end{align}
In HBCHPT we project on velocity-dependent fields $\mathcal{N}$ via
\begin{align}
\label{nucleonprojection}
\NN=(1/2)(1+v_\mu\gamma^\mu)\exp(i mv\cdot x)\Psi\,,
\end{align}
with $v$ a
four-velocity with $v^2=1$. However, in this paper we need to introduce also an
antinucleon field with the same velocity $v$. The charge conjugate
fermion spinor is $\psi^c\equiv  -i\gamma^2\psi^*$. We then define
\begin{align}
\Psi^c \equiv\,& i\tau^2 \left(\begin{array}{c}p^c\\n^c\end{array}\right) = \left(\begin{array}{c}n^c\\-p^c\end{array}\right)\,,
& \Psi^c\to\,&h \Psi^c\,.
\end{align}
The transformation under the chiral group follows from the properties
of $SU(2)$ using the identities in App.~\ref{app:group}. We then define a HBCHPT field for the antineutron as
\begin{align}
\NNc=(1/2)(1+v_\mu\gamma^\mu)\exp(i mv\cdot x)\Psi^c\,.
\end{align}
Compared to the first projection (\ref{nucleonprojection}), this is at $-v$
if formulated in terms of $\Psi$. $\mathcal{N}$ and $\mathcal{N}^c$ are in
HBCHPT independent fields, since they are from expansions around
different widely-separated velocities as
depicted in Fig.~\ref{fig:velocities}.
The lowest order Lagrangian for the HBCHPT fields is
\begin{align}
\label{LOHBCHPT}
\mathcal{L}_{N} =\,&
\overline\NN\left(i v^\mu D_\mu + g_A u^\mu S_\mu\right)\NN
+
\overline\NNc\left(i v^\mu D_\mu - g_A u^\mu S_\mu\right)\NNc\,.
\end{align}
The signs can be derived using charge conjugation. The spin vector
$S_\mu$ has the properties
\begin{align}
S_\mu=\,&-\frac{1}{4}\gamma_5\left[\gamma_\mu,\gamma_\nu\right]v^\nu\,,
& S^2 =\,& \frac{1-d}{4}\,,
& \left\{S_\mu,S_\nu\right\} =\,& \frac{1}{2}\left(v_\mu v_\nu-g_{\mu\nu}\right)\,,
& v\cdot S =\,&0\,.
\end{align}
These properties are sufficient for our calculation.
Higher order Lagrangians can be constructed in the same way as usual. 

The operators that give neutron-antineutron transitions have to be written with
doublet indices and must create the antineutron. For this we introduce\footnote{In this equation, the fields $p^c,n^c$ are the HBCHPT ones for the antinucleons, not the relativistic fields.}
\begin{align}
\NNt =\,&\left(\begin{array}{c}\overline{p^c}\\\overline{n^c}\end{array}\right)
= -i\tau^2{\overline\NNc}^T\,,
& \NNt\to\,& h\NNt\,.
\end{align}
We need to construct operators that transform with left- or right-handed
doublet indices under $SU(2)_L\times SU(2)_R$. These can then be contracted
with the spurion operators given in Tab.~\ref{tab:operators} to make invariant
quantities.

To be precise, a lower index on an object $x_{i_L}$ leads to the transformation
$x_{i_L}\to \sum_{j_L} (g_L)_{i_L}^{~ j_L}x_{j_L}$  and equivalently for a
right-handed lower index. Some examples of objects with the corresponding
indices are:
\begin{align}
\left(Ui\tau^2\right)_{i_R j_L},~~~\left(u\NN\right)_{i_R},~~~
\left(u^\dagger\NN\right)_{i_L},~~~\left(u\NNt\right)_{i_R},~~~
\left(u^\dagger\NNt\right)_{i_L},~~~\left(u^\dagger u_\mu ui\tau_2\right)_{i_L j_L}\,.
\end{align}
To get a neutron to antineutron transition we need an $\NNt$ and a $\NN$ field.
Dirac (or fermion) indices are contracted between these.

The lowest order, $p^0$, operators are
\begin{align}
\label{LOoperators}
(3_L,1_R) :\,& R_{i_L j_L}=\left(u^\dagger\NNt\right)_{i_L} \left(u^\dagger\NN\right)_{j_L}
\nonumber\\
(3_L,5_R) :\,&R_{i_L j_L k_R l_R m_R n_R}=\left(u\NNt\right)_{k_R} \left(u\NN\right)_{l_R}
 \left(Ui\tau^2\right)_{m_R i_L}
 \left(Ui\tau^2\right)_{n_R j_L}
\nonumber\\
(7_L,1_R) :\,&\text{---}
\end{align}
and the parity-conjugates. There is no lowest order operator for $(7_L,1_R)$.
The first operator that appears for $(7_L,1_R)$ is at order $p^2$:
\begin{align}
(7_L,1_R), p^2:\,&
\left(u^\dagger\NNt\right)_{i_L} \left(u^\dagger\NN\right)_{j_L}
\left(u^\dagger u_\mu ui\tau_2\right)_{k_L l_L}\left(u^\dagger u_\mu ui\tau_2\right)_{m_L n_L}
\end{align}
At higher orders there are very many operators. A partial list can be found
in \cite{masterthesis}. We will restrict ourselves to comments sufficient for
the application to neutron-antineutron transitions. The relevant independent
combinations we refer to as $\delta_i$ below.

At order $p$, the operators must contain a derivative $D_\mu$ or $u_\mu$.
As such, they will contain either dependence on the neutron or antineutron
four momentum,
or contain an extra pion. For a neutron-antineutron transition at rest
the HBCHPT momentum $k_\mu$ vanishes. There is thus no tree level contribution
to neutron-antineutron transitions. Loop level contributions from these
operators will start at $p^3$, which is beyond what is considered in this paper.

At order $p^2$ there are very many operators that contribute,
a rather extensive list is in \cite{masterthesis}. Two examples are
\begin{align}
 \left(u^\dagger D_\mu\NNt\right)_{i_L} \left(u^\dagger D^\mu\NN\right)_{j_L},~~~
 \left(u^\dagger\NNt\right)_{i_L} \left(\chi^\dagger u\NN\right)_{j_L}\,.
\end{align}
For this paper it is sufficient to notice that there is a free parameter
at order $p^2$ associated with each operator.

How many parameters do we need to order $p^2$ to describe
neutron-antineutron transitions given the operators $P_1,\ldots,P_7$
with a given coefficient? The operators $P_1,P_2,P_3$ are all $(3_L,1_R)$,
however the quark-operators are not related by a chiral transformation.
This leads to three free parameters at order $p^0$ and three more at order
$p^2$. The three operators $P_5,P_6,P_7$ belong to same chiral multiplet, i.e.
they are related via a chiral transformation. This leads to one parameter
at $p^0$ and one more at $p^2$. The $(7_L,1_R)$ operator at order $p^2$ does not
contribute to neutron-antineutron transitions.

The values to which the spurions need to be set to reproduce the quark level
operators can be derived from the expressions in \cite{Buchoff:2015qwa}.
They are ($1$ corresponds to an up-quark, $2$ to a down-quark):
\begin{align}
\label{spurionvalues}
\theta_1^{ij} =\,&\theta_2^{ij} =\theta_3^{ij} = \delta^i_2 \delta^j_2\,,
\nonumber\\
\theta_5^{ijklmn}=\,&\delta^i_1\delta^j_1\delta^k_2\delta^l_2\delta^m_2\delta^n_2\,,
\nonumber\\
\theta_6^{ijklmn}=\,&\frac{1}{2\sqrt2}
\left(\delta^i_1\delta^j_2+\delta^i_2\delta^j_1\right)
\left(\delta^k_1\delta^l_2\delta^m_2\delta^n_2
     +\delta^k_2\delta^l_1\delta^m_2\delta^n_2
     +\delta^k_2\delta^l_2\delta^m_1\delta^n_2
     +\delta^k_2\delta^l_2\delta^m_2\delta^n_1\right)\,,
\nonumber\\
\theta_7^{ijklmn}=\,&\frac{1}{\sqrt6}
\delta^i_2\delta^j_2
\left(\delta^k_1\delta^l_1\delta^m_2\delta^n_2
     +\delta^k_1\delta^l_2\delta^m_1\delta^n_2
     +\delta^k_1\delta^l_2\delta^m_2\delta^n_1
     +\delta^k_2\delta^l_1\delta^m_1\delta^n_2
     +\delta^k_2\delta^l_1\delta^m_2\delta^n_1
     +\delta^k_2\delta^l_2\delta^m_1\delta^n_1\right)\,.
\end{align}
Note that these are normalized to 1, slightly different from~\cite{Buchoff:2015qwa}.

To summarize the neutron-antineutron part.
If the Lagrangian at the quark-level is of the form
\begin{align}
\sum_{i=1,7} \alpha_i P_i
\end{align}
then the LO ChPT Lagrangian has the form
\begin{align}
\label{Lagrangiannnbar}
\mathcal{L}_{n\bar n} =\,&
\left(\beta_1\alpha_1+\beta_2\alpha_2+\beta_3\alpha_3\right)\theta_1^{i_Lj_L}R_{i_Lj_L}
\nonumber\\&
+\beta_5\left(\alpha_5\theta_5^{i_Lj_Lk_Rl_Rm_Rn_R}+\alpha_6\theta_6^{i_Lj_Lk_Rl_Rm_Rn_R}+\alpha_7\theta_7^{i_Lj_Lk_Rl_Rm_Rn_R}\right)R_{i_Lj_Lk_Rl_Rm_Rn_R}
\end{align}
with the spurions as defined in (\ref{spurionvalues}) and the operators
in (\ref{LOoperators}). The $\alpha_i$ are short-distance parameters while the
$\beta_i$ are long-distance parameters.
The parity-conjugate operators can be included similarly.

\section{Analytical results}
\label{sec:analytical}

\begin{figure}
\centerline{
\includegraphics[width=0.8\textwidth]{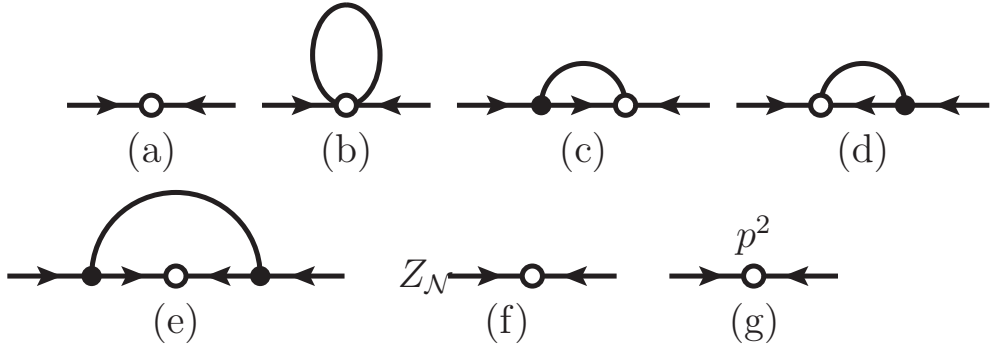}}
\caption{\label{figdiagrams}The diagrams for $n\bar n$ transitions to
order $p^2$. An open dot indicates a vertex from the $n\bar n$ Lagrangian
(\ref{Lagrangiannnbar}), a dot from the LO normal Lagrangian
(\ref{LOHBCHPT}). The contributions from wave-function renormalization are
indicated schematically in(f) and from the $p^2$ $n\bar n$-Lagrangian in (g).
A right-pointing line is a neutron, a left-pointing line an antineutron.}
\end{figure}

The diagrams needed for $n\bar n$ transition to order $p^2$ are shown
in Fig.~\ref{figdiagrams}.

The LO, $p^0$, result from Fig.~\ref{figdiagrams}(a) is
\begin{align}
A(n\to\bar n)_{LO} =\,& \beta_1\alpha_1+\beta_2\alpha_2+\beta_3\alpha_3
 +\beta_5\left(\alpha_5-\frac{\alpha_6}{\sqrt2}+\frac{\alpha_7}{\sqrt6}\right)\,.
\end{align}

The integrals we use are defined in App.~\ref{app:integrals}. 
The tadpole diagram of Fig.~\ref{figdiagrams}(b) contributes
\begin{align}
A(n\to\bar n)_{(b)} =\,&\frac{1}{F^2} A(m_\pi^2)\Bigg[\left(\beta_1\alpha_1+\beta_2\alpha_2+\beta_3\alpha_3\right)
 +7\beta_5\left(\alpha_5-\frac{\alpha_6}{\sqrt2}+\frac{\alpha_7}{\sqrt6}\right)
\Bigg]\,.
\end{align}

The diagrams (c) and (d) contain the integral
\begin{align}
\frac{1}{i}\int \frac{d^d r}{(2\pi)^d}\frac{S\cdot r}{(r^2-m_\pi^2)(v\cdot (r+k))}\,.
\end{align}
We work in the frame where the external momentum $k$ vanishes. In infinite
volume the integral is proportional to $v\cdot S=0$. In finite volume for a
neutron and antineutron at rest, $S$ is purely spatial, and the integral/sum is
odd under $\vec r\to -\vec r$ and vanishes for periodic boundary
conditions. So (c) and (d) give no contribution.

Diagram (e) can be rewritten in terms of the integral
\begin{align}
I(m_\pi^2) = \frac{1}{i}\int \frac{d^dr}{(2\pi)^2}\frac{\left(S\cdot r\right)^2}{(r^2-m_\pi^2) (v\cdot r)^2}\,.
\end{align}
The central vertex is directly the LO contribution so (f) contributes
\begin{align}
A(n\to\bar n)_{(f)} =\,&-\frac{g_A^2}{F^2}I(m_\pi^2) A(n\to\bar n)_{LO}\,.
\end{align}

Wave-function renormalization can be computed from the derivative of the
nucleon (and antinucleon) selfenergy. This leads again to the occurrence of
the integral $I(m_\pi)^2$ in this contribution. We get
\begin{align}
A(n\to\bar n)_{(f)} =\,&\frac{3g_A^2}{F^2}I(m_\pi^2) A(n\to\bar n)_{LO}\,.
\end{align}
Depending on the form of $p^3$ Lagrangian in the pion nucleon sector
chosen, we have a contribution proportional to $m_\pi^2$ and a possible
$p^3$ pion-nucleon LEC. This is nonzero if choosing the Lagrangian
in~\cite{Bernard:1995dp} and vanishes if the version of~\cite{Ecker:1995rk}
is chosen. The two choices are related by a field redefinition.
The effect is that the $p^2$ $n\bar n$ LECs (referred to as $\delta_i$ below)
have different values in the two cases but such that the total result remains
the same.

The final result is
\begin{align}
\label{finalresult}
A(n\to \bar n) =\,&
\left(\beta_1\alpha_1+\beta_2\alpha_2+\beta_3\alpha_3\right)
\left[1+\frac{1}{F^2}\left(A(m_\pi^2)+2g_A^2 I(m_\pi^2)\right)\right]
\nonumber\\&
+\beta_5\left(\alpha_5-\frac{\alpha_6}{\sqrt2}+\frac{\alpha_7}{\sqrt6}\right)
\left[1+\frac{1}{F^2}\left(7A(m_\pi^2)+2g_A^2 I(m_\pi^2)\right)\right]
\nonumber\\&
+m_\pi^2\left(\delta_1\alpha_1+\delta_2\alpha_2+\delta_3\alpha_3\right)
+m_\pi^2\delta_5\left(\alpha_5-\frac{\alpha_6}{\sqrt2}+\frac{\alpha_7}{\sqrt6}\right)\,.
\end{align}
In order to get the infinite volume finite result,
replace the $\delta_i$ by their finite parts $\delta_i^r$ and the
integrals $I,A$ by $\overline I, \overline A$. The finite volume correction
is obtained by dropping terms not involving an integral and replacing
$I,V$ by $I^V,A^V$. Expressions for these integrals are in
App.~\ref{app:integrals}.

\section{Numerical results}
\label{sec:numerical}

We set in this section all $p^2$ LECs, $\delta_i^r$, to zero.

\begin{figure}
\begin{minipage}{0.49\textwidth}
\includegraphics[width=0.99\textwidth]{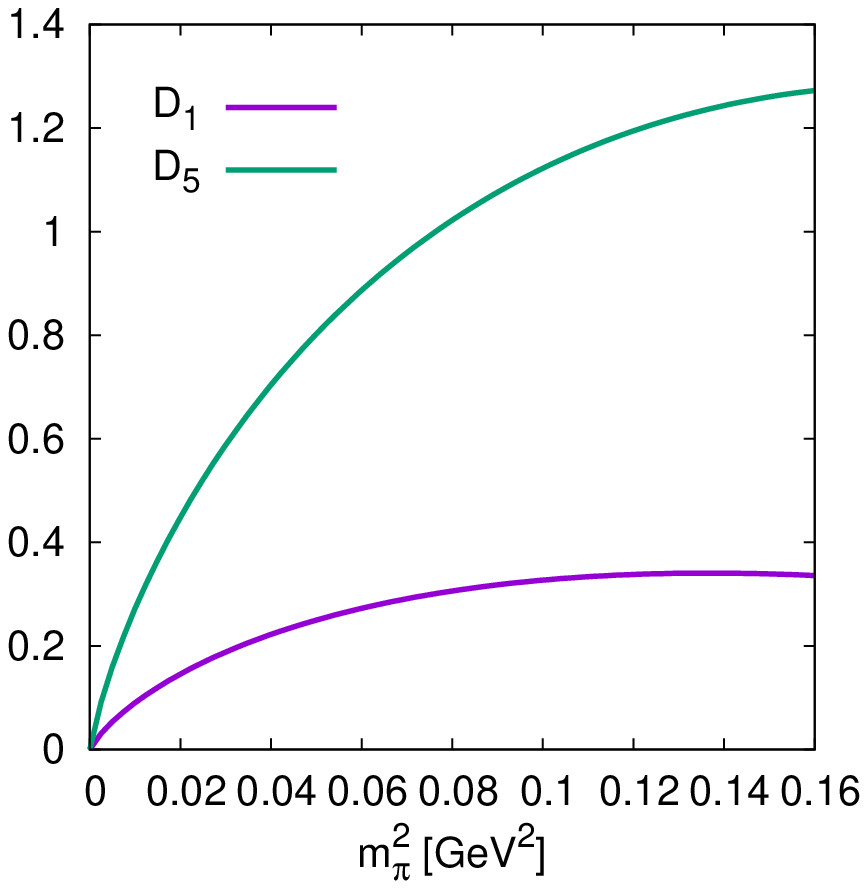}
\centerline{(a)}
\end{minipage}
\begin{minipage}{0.49\textwidth}
\includegraphics[width=0.99\textwidth]{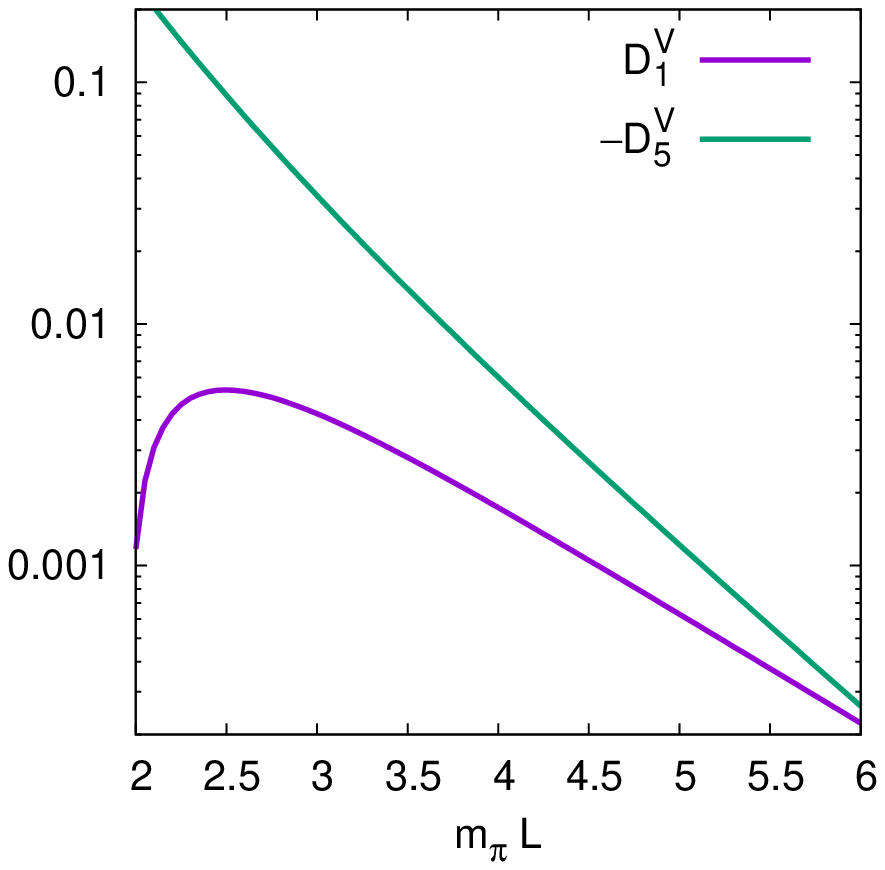}
\centerline{(b)}
\end{minipage}
\caption{\label{fignumerics} The numerical results of the pure loop contributions. (a) The infinite volume correction of (\ref{correctioninfinite})
(b) The finite volume correction of (\ref{correctionfinite}).}
\end{figure}

The relative chiral correction from the loops to $(3_L,1_R)$ ($D_1)$) and
$(3_L,5_R)$ ($D_5$) operators is given by keeping the $I,A$ terms
in (\ref{finalresult})
and replacing them by $\overline I,\overline A$. The result is
\begin{align}
\label{correctioninfinite}
D_1 =\,& \frac{m_\pi^2}{16\pi^2 F^2}\left[\left(-1-\frac{3g_A^2}{2}\right)
  \log\frac{m_\pi^2}{\mu^2}-g_A^2\right]\,,
\nonumber\\
D_5 =\,& \frac{m_\pi^2}{16\pi^2 F^2}\left[\left(-7-\frac{3g_A^2}{2}\right)
  \log\frac{m_\pi^2}{\mu^2}-g_A^2\right]\,.
\end{align}
These are plotted in Fig.~\ref{fignumerics}(a) for a range of $m_\pi^2$
with $F=92.2$~MeV fixed and $g_A=1.25$. Note that they are large for the
$(3_L,5_R)$ operators alread at $m_\pi\approx200$~MeV.

The correction due to finite volume is obtained by replacing
$I,A$ by $I^V,A^V$ in (\ref{finalresult}):
\begin{align}
\label{correctionfinite}
D_1^V =\,& \frac{1}{F^2}\left[\left(1+\frac{g_A^2}{2}\right)A^V(m_\pi^2,1)
         +m_\pi^2 g_A^2 A^V(m_\pi^2,2)\right]\,,
\nonumber\\
D_5^V =\,& \frac{1}{F^2}\left[\left(7+\frac{g_A^2}{2}\right)A^V(m_\pi^2,1)
         +m_\pi^2 g_A^2 A^V(m_\pi^2,2)\right]\,.
\end{align}
These are plotted in Fig.~\ref{fignumerics}(b) for $m_\pi=135$~MeV and
$F=92.2$~MeV as a function of $m_\pi L$. $D_5^V$ is negative over the whole
region while $D_1^V$ is positive. $D_1^V$ goes through zero just below the
region plotted. The finite volume corrections are small for $m_\pi L>4$.

\section{Conclusions}
\label{sec:conclusions}

In this paper we have constructed ChPT operators for the dimension 9 six-quark
operators that contribute to neutron-antineutron oscillations. At order $p^0$
there is one term each transforming as $(3_L,1_R)$ and $(3_L,5_R)$.
The $(7_L,1_R)$ operators only contribute at order $p^3$ by power-counting but
do require isospin violation. We showed that the order $p$ operators only
contribute from order $p^3$. There is a large number of operators
contributing at order $p^2$, a partially complete list can be found
in \cite{masterthesis}. The same is true for the parity-conjugate operators.

Our main results are the one-loop corrections
in (\ref{finalresult}), (\ref{correctioninfinite})
and (\ref{correctionfinite}). We have shown numerical results.
The finite volume corrections are small for $m_\pi L > 4$. We found that
chiral corrections are reasonable for the $(3_L,1_R)$ operators but can be
sizable for the $(3_L,5_R)$ operators.

\section*{Acknowledgements}

This work is supported in part by the Swedish Research Council grants
contract numbers 621-2013-4287, 2015-04089 and 2016-05996 and by
the European Research Council (ERC) under the European Union's Horizon 2020
research and innovation programme (grant agreement No 668679).

\appendix
\section{Group theory}
\label{app:group}

$SU(2)$ is a pseudoreal group with as generators $T^a=(1/2)\tau^a$. The Pauli
matrices $\tau^a$ are Hermitian and satisfy
\begin{align}
\tau^{aT} = \tau^{a*} = -\tau^2\tau^a\tau^2\,.
\end{align}
As a consequence the special unitary matrices $x=g_L,g_R,u,h$ all satisfy
\begin{align}
\tau^2 x \tau^2 =\,& x^*\,, & \tau^2 x^T \tau^2 = x^\dagger\,.
\end{align}
These identities are used a lot in the construction of the transformations
and operators in the main text.

\section{Integrals}
\label{app:integrals}

The integrals we need to calculate both at infinite and finite volume.
In finite volume we replace the integral over spatial momenta by a sum.
The techniques are well known both at finite and infinite volume, we use
here \cite{Bernard:1995dp} and \cite{Bijnens:2014yya}.

The mesonic integral/sum needed is
\begin{align}
A(m^2)=\,&\frac{1}{i}\int \frac{d^dr}{(2\pi)^d}\frac{1}{r^2-m_\pi^2}
= \frac{\lambda_0}{16\pi^2}+\overline A(m^2)+A^V(m^2,1)\,, 
\end{align}
with
\begin{align}
\lambda_0 =\,& \frac{1}{\epsilon}+1+\log(4\pi)-\gamma_E\,, & d=\,&4-2\epsilon\,.
\end{align}
The terms with $\lambda_0$ are removed by the renormalization procedure
and the logarithms of $m_\pi^2$ obtain the subtraction scale $\mu^2$ via the
renormalization as well. We therefore quote the integrals including $\mu^2$.
The finite volume part depends on the spatial length scale $L$.
The results are, see e.g. \cite{Bijnens:2014yya},
\begin{align}
\overline A(m_\pi^2) =\,&-\frac{m_\pi^2}{16\pi^2}\log\frac{m_\pi^2}{\mu^2}\,,
\nonumber\\
A^V(m^2,n) =\,&\frac{(-1)^n}{16\pi^2}\left(\frac{L^2}{4}\right)^{n-2}
\int\frac{d\lambda}{\Gamma(n)}\lambda^{n-3}e^{-\lambda m^2 L^2/4}
\left[\theta_3\left(e^{-1/\lambda}\right)^3-1\right]\,,
\nonumber\\
\theta_3 =\,&\sum_{n=-\infty,\infty} q^{(n^2)}\,.
\end{align}

The other integral/sum needed is
\begin{align}
I(m_\pi^2) = \frac{1}{i}\int \frac{d^dr}{(2\pi)^d}
\frac{\left(S\cdot r\right)^2}{(r^2-m_\pi^2)(v\cdot r)^2}\,.
\end{align} 
The numerator can be rewritten via
\begin{align}
\left(S\cdot r\right)^2 = \frac{1}{2} r^\mu r_\nu\left\{S_\mu,S_\nu\right\}
 = \frac{1}{4}\left[(v\cdot r)^2-r^2\right]
 =\frac{1}{4}\left[(v\cdot r)^2-(r^2-m_\pi^2)-m_\pi^2\right] \,.
\end{align}
The second term leads to an integral with only $v\cdot r$ in the denominator.
These vanish both at infinite and finite volume. We thus get
\begin{align}
I(m_\pi^2) = \frac{1}{4}A(m_\pi^2)-\frac{m_\pi^2}{4}
\frac{1}{i}\int \frac{d^dr}{(2\pi)^d}
\frac{1}{(r^2-m_\pi^2)(v\cdot r)^2}
\end{align}
We combine the propagators in the second term with
$1/(ab^2)=\int_0^\infty d\lambda\, 8\lambda/(a+2b\lambda)^3$ and shift the momentum
to\footnote{This also works at finite volume since $v$ is in the temporal direction there.} $\tilde r=r+v\lambda$ and obtain the integral
\begin{align}
\frac{1}{i}\int \frac{d^d\tilde r}{(2\pi)^d}
\int_0^\infty d\lambda 8\lambda\frac{1}{\left(r^2-(m_\pi^2+\lambda^2)\right)^3}
= -\frac{1}{i}\int \frac{d^d\tilde r}{(2\pi)^d}\frac{2}{\left(\tilde r^2-m_\pi^2\right)^2}\,.
\end{align}
In the second step we have done the $\lambda$-integral. The integral/sum
appearing now is known and we obtain
\begin{align}
I(m_\pi^2) =\,& \frac{\lambda_0}{16\pi^2}\frac{3m_\pi^2}{4}+\overline I(m_\pi^2)
+I^V(m_\pi^2)\,,
\nonumber\\
\overline I(m_\pi^2) =\,& \frac{m_\pi^2}{16\pi^2}\left(-\frac{3}{4}\log\frac{m_\pi^2}{\mu^2}-\frac{1}{2}\right)\,,
\nonumber\\
I^V(m_\pi^2) =\,&\frac{1}{4}A^V(m_\pi^2,1)+\frac{m_\pi^2}{2}A^V(m_\pi^2,2)\,.
\end{align}

\end{document}